\documentclass[aps,pra,twoside,twocolumn,showpacs,floatfix,pdflatex]{revtex4-1}

\usepackage{latexsym}
\usepackage{graphicx}
\usepackage{amsmath, amssymb}
\usepackage{color}

\newcommand{\ket}[1]{|#1 \rangle}
\newcommand{\bra}[1]{\langle #1|}

\newcommand{\ketbra}[1]{\ket{#1}\!\bra{#1}}
\DeclareMathOperator{\Tr}{Tr}
\DeclareMathOperator{\sinc}{sinc}

\begin{document}

\title{Qubit carriers with internal degrees of freedom in a non-factorable state}

\author{Martina Mikov\'{a}}
\author{Helena Fikerov\'{a}}
\author{Ivo Straka}
\author{Michal Mi{\v{c}}uda}
\author{Miroslav Je\v{z}ek}
\author{Miloslav Du{\v{s}}ek}
\author{Radim Filip}

\affiliation{Department of Optics, Faculty of Science, Palacky University,
             17.~listopadu 12, 771\,46 Olomouc, Czech Republic}

\begin{abstract}
A directly measurable parameter quantifying {\em effective indistinguishability} of particles as a resource for quantum information transfer and processing is proposed. In contrast to commonly used overlap of quantum states of particles, defined only for a factorable states, this measure can be generally applied to any joint state of the particles. The relevance of this generalized measure for photons produced in parametric down-conversion has been experimentally verified. The simplest linear-optical quantum-state-transfer protocol, for which this measure directly determines fidelity of the transferred state, was experimentally tested. It has been found that even if some degrees of freedom of two particles are entangled, the particles can still serve as good carriers of qubits.
\end{abstract}

\maketitle

\section{Introduction}

In last few decades quantum physics has offered novel applications in information and communication technology. Their performance crucially depends on the quality of elements of quantum information -- qubits \cite{NielsenChuang}. Necessary conditions for high-fidelity qubits are their coherence \cite{MandelWolf,Zurek2003}, which appears when no information is leaking into an environment, and indistinguishability \cite{Peres,HOM,Santori2002}. By {\em effective} indistinguishability of two (spatially separated) particles we mean that all their internal degrees of freedom not used to carry information are identical. For factorable state $\rho_A \otimes \rho_B$ of two particles $A$ and $B$ it means that the action of a flip (exchange) operator, $F \left( \rho_A \otimes \rho_B \right) F = \rho_B \otimes \rho_A$, does not change the state.
The mean value of the flip operator for a factorable state equals to the overlap of states of individual subsystems, $\Tr \left[ F (\rho_A\otimes\rho_B) \right] = \Tr \left[ \rho_A\rho_B \right]$ \cite{Werner1989}. This is a hint for an indistinguishability measure. But these considerations are still valid {\em only} for factorable states.
Direct measurement of the overlap was already suggested for qubits \cite{Ekert2002} and harmonic oscillators \cite{Filip2002}. For the simplest case of two photons, it was measured by Hong-Ou-Mandel (HOM) type interferometry \cite{Hendrych2003}. There are many other related two-photon experiments \cite{beu06,hof12,tho06,mau07,pat10,ben09,boi09}. Indistinguishability of particles is crucial in a number of quantum protocols which were intensively studied in recent years and are still of a great interest \cite{wan06,hal07,nag09,bri03,lem11,mic08,cer08,kac10,per10,par07,cre11}.

In this paper, we propose a directly measurable parameter, $D$, quantifying effective indistinguishability of particles which can be used for an arbitrary state. Effective indistinguishability can be defined by means of the flip operator exchanging relevant degrees of freedom of the particles. Full flip of particles corresponds to complete exchange of their quantum states. On the other hand, transfer by quantum teleportation relies both on the particle indistinguishability and entanglement. Therefore, to show how distinguishability of particles used as information carriers affects quantum information processing without the influence of other imperfections of resources, we designed a quantum-state-transfer protocol depending uniquely on indistinguishability of particles. We consider transfer of a state of a source single-photon qubit ($S$) to a single-photon target qubit ($T$). The transfer is performed by a partial exchange of photons, optimal measurement on $S$, and conditional feed-forward correction on $T$. We show that fidelity of the transferred state depends directly on $D$ even if the internal degrees of freedom of the particles are entangled.
In more complex quantum  protocols, the quality of information processing may depend on a nontrivial combination of the effective indistinguishability and the properties of other resources.

\section{Operational measure of indistinguishability}

Let us have two particles, $S$ and $T$, carrying the same qubit states, let $\rho_{E,ST}$ denote the state, not necessarily separable, of all of the other (inaccessible) degrees of freedom. The internal \emph{environmental} degrees of freedom can even be entangled with an external environment. Clearly, they are responsible for distinguishability of the particles. Let us define a measure $|D|$ describing an effective indistinguishability, where
\begin{equation}\label{D-def}
 D = \Tr \left[ F_A \, \rho_{E,ST} \right].
\end{equation}
Operator $F_A=\sum_{m,n}|\psi_n\rangle_S\langle\psi_m|\otimes |\psi_m\rangle_T\langle\psi_n|$ is a flip operator acting on the joint environment of both particles, which exchange basis states corresponding to a given observable $A$, where $A|\psi_n\rangle=a_n|\psi_n\rangle$. Properties of $\Tr \left[ F_A \, \rho_{E,ST} \right]$ follows from the features of operator $F_A$. Operator $F_A$ is both hermitian $F_A=F_A^{\dagger}$ and unitary $F_AF^{\dagger}_A=F^{\dagger}_AF_A=1$ \cite{Werner1989}. Since $F_A$ is commuting with any local unitary transformation $U_{ES}\otimes U_{ET}$, it is invariant to a choice of operator $A$ and therefore, we can consider $F$ instead $F_A$ as a basis-independent operation. $F$ can be also expressed as a difference, $F=\Pi_\mathrm{sym} - \Pi_\mathrm{anti}$, of orthogonal projectors $\Pi_\mathrm{sym}$ and $\Pi_\mathrm{anti}$ onto the symmetric and anti-symmetric subspace of the total space, respectively. Therefore it is directly linked to indistinguishability of the environmental states. Clearly, $F$ is a dichotomic observable (with eigenvalues $\pm 1$) and we get $-1 \leq \Tr \left[ F \rho \right] \leq 1$.
Parameter $D$ is invariant under symmetric local unitary transformations $U_{ES}\otimes U_{ET}$ (indistinguishability cannot change if the particles go through the same unitary channels). In contrast, entanglement of the environments have to be invariant under more general unitary transformations $U_{ES}\otimes V_{ET}$, where $U_{ES}$ may differ from $V_{ET}$. The conceptual difference is that a local unitary applied on the environment of one particle can make its state distinguishable from the state of the other particle, although it does not change the amount of correlations and entanglement between the environments. Any symmetric state of two particles satisfying $\rho_{E,ST} = F \rho_{E,ST}=\rho_{E,ST} F$ has $D=1$, irrespective of its entanglement. States related by the permutation operation, $\rho'_{E,ST} = F \rho_{E,ST} F$, have the same values of $D$. For separable state $\rho_{E,ST} = \sum_{n} p_{n} \, \rho_{n,ES} \otimes \rho_{n,ET}$ of two particles, $D$ is always positive semi-definite. If a twirling transformation is applied to any input state $\rho_{E,ST}$  (i.e., if identical random unitaries are applied to both qubits) parameter $D$ does not change and the resulting state is the Werner state which is fully parameterized by $D$. For factorable states $D = \Tr \left[ \rho_{ES} \rho_{ET} \right]$ and it reduces to the overlap of $S$ and $T$ particles.
Notice however, that in general $D$ is not equal to the overlap. The problem of the overlap lies in the assumption that quantum states of all, even unaccessible, degrees of freedom of two bosons are factorized. Such assumption cannot be operationally certified, except one achieves an unrealistic complete tomography of a joint quantum state corresponding to all degrees of freedom of both particles.

Alternatively, the flip operator can be rewritten as $F = \left( | \tilde{\Psi}_+ \rangle_{ST} \langle \tilde{\Psi}_+ | \right)^{T_A}$ representing an entanglement witness, where $|\tilde{\Psi}_+ \rangle=\sum_{k}|\psi_k\rangle_S |\psi_k\rangle_T$ is an unnormalized symmetric state. Thus $D<0$ is a witness of entanglement in $\rho_{E,ST}$ (due to the presence of an anti-symmetric component) \cite{Werner1989,Filip2002,ent_wit_exp}. However, we do not focus on entanglement. Instead, indistinguishability is in our attention.

Measure $|D|$ characterizes resources, i.e. quantum
systems, information is encoded in. Among other conditions,
theory of quantum information processing requires all the
resources to be in the same states which are decoupled from
each other (their total state must be factorable). In our
notation $\rho_{E,ST} = \rho_{E,S} \otimes \rho_{E,T}$ with
$\rho_{E,S} = \rho_{E,T}$. Then they can be used to represent
the ideal qubits (or qudits). But this strict condition is not
always fulfilled in practice. In case of linear optical quantum
information processing, $|D|=1$ says that the resources behave
in the same way as if they fulfilled the upper condition even
if they actually do not. It means, they can be used for
encoding and processing qubits (qudits) even if some of their
degrees of freedom are, e.g., entangled. Interestingly, this is
exactly the case of traditional photon pairs generated by SPDC. Information
is usually encoded into polarization or spatial degrees of
freedom but frequency degrees of freedom are entangled. Thus
parameter $|D|$ quantifies effective indistinguishability of
resources for quantum information processing.

\section{Photonic qubit transfer}

To demonstrate relevance of effective indistinguishability $|D|$, we have proposed and experimentally tested the simplest example of a quantum information transfer, in which $|D|$ alone directly determines quantum fidelity of the transferred states. It manifests a clear operational meaning of the above defined effective indistinguishability.

We consider only the equatorial states of qubit $S$ represented by a dual-rail superposition of single photon states
\begin{equation}\label{state_S}
  |\Psi\rangle_S=\frac{1}{\sqrt{2}}[|0,1\rangle_S+\exp(i\theta)|1,0\rangle_S],
\end{equation}
where phase $\theta$ may be unknown during the transfer. This state should be transferred to target qubit $T$ represented by another single photon, which is in state
\begin{equation}\label{state_T}
  |\Phi\rangle_T=\frac{1}{\sqrt{2}}(|0,1\rangle_T+|1,0\rangle_T)
\end{equation}
at the beginning. All other degrees of freedom are described by a density matrix $\rho_{E,ST} = \sum_{i,j,k,l} c_{ij,kl} |\psi_i \rangle_{ES} \langle \psi_j| \otimes | \psi_k \rangle_{ET} \langle \psi_l |$, where $i,j,k,l$ are multi-indices over many different sub-degrees of freedom of the joint ``environment'' of photons and $| \psi_x \rangle$ denote basis states of each photon. We consider that all physical differences between the particles are contained in this environmental state.
So the overall initial state of the qubits reads $\rho_\mathrm{ini} = | \Psi \rangle_S\langle \Psi | \otimes | \Phi \rangle_T \langle \Phi | \otimes \rho_{E,ST}$.

\begin{figure}
  \begin{center}
   %\smallskip
  \resizebox{\hsize}{!}{\includegraphics*{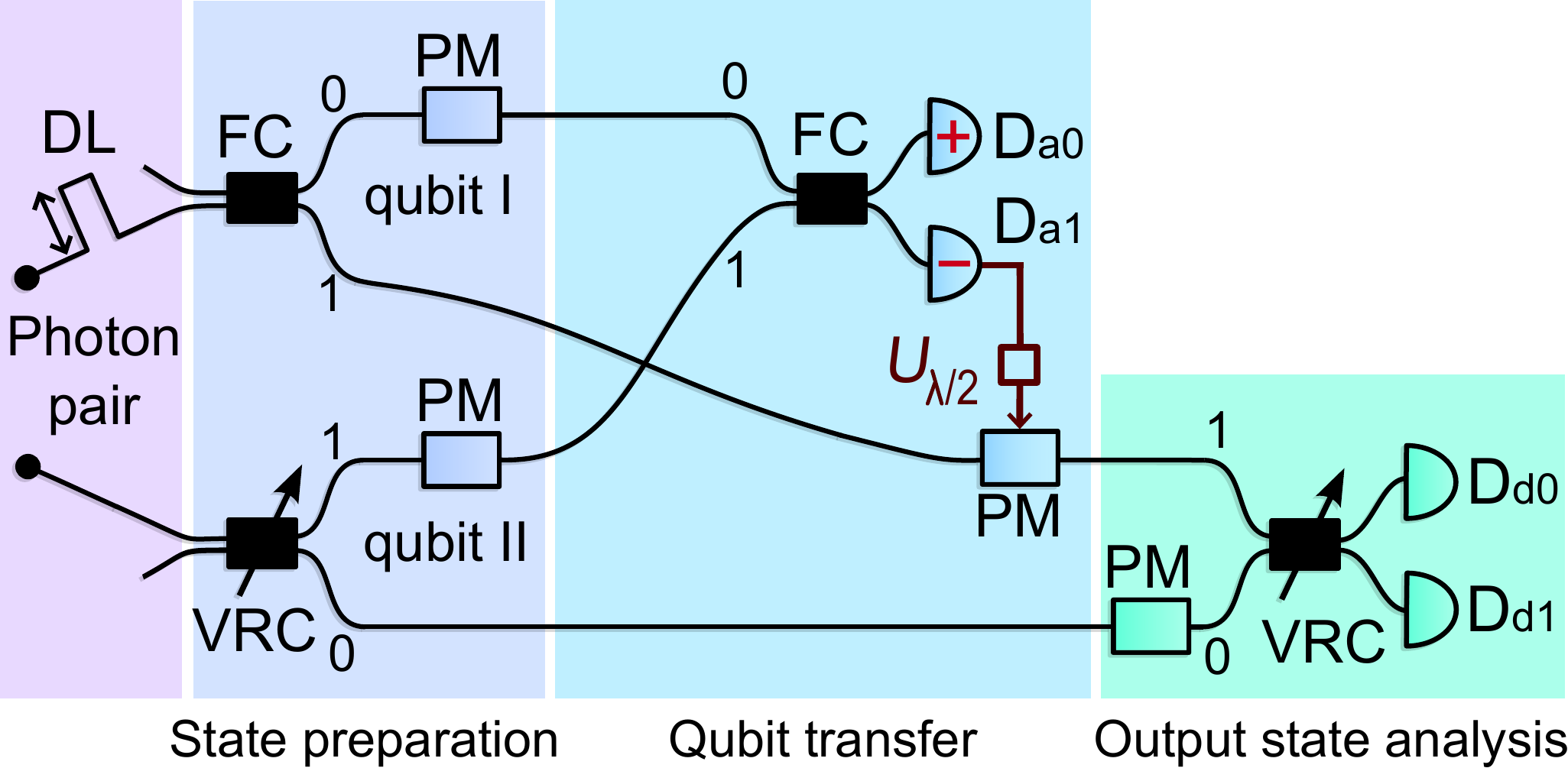}}
   %\smallskip
  \end{center}
  \caption{(Color online) Scheme of the experiment. FC -- fiber couplers,
           VRC -- variable ratio couplers, PM -- phase modulators, DL -- delay line, D -- detectors. The couplers and phase modulators in the \emph{State preparation} stage enable us to prepare required qubit states (each qubit is represented by a single photon which may propagate in two optical fibers). They do not affect the environmental degrees of freedom. The middle PM applies conditional phase shift depending on the result of the auxiliary measurement. It is a part of the protocol. The rightmost PM and VRC serve for output state tomography.}
  \label{fig-scheme}
\end{figure}

In general, an imperfect interaction between qubits can also limit quality of the transfer. We therefore consider implementation without any direct interaction. To transfer a quantum state, we swap two rails between $S$ and $T$ (see Fig.~\ref{fig-scheme}). It means, basis states $|0,1\rangle_S|1,0\rangle_T|\psi_i\rangle_{ES}|\psi_k\rangle_{ET}$ are changed to $|0,1\rangle_S|1,0\rangle_T|\psi_k\rangle_{ES}|\psi_i\rangle_{ET}$ (here the environmental basis states are swapped). While basis states $|1,0\rangle_S|0,1\rangle_T|\psi_i\rangle_{ES}|\psi_k\rangle_{ET}$ remain unchanged. All other possibilities are excluded by post-selection considering only single photon in $S$ and single photon in $T$. This conditional ``partial exchange'' process entangles qubits together with the basis states of the environment, resulting in the following total state of two photons: $\sum_{i,j,k,l} c_{ij,kl} |\Psi_{i,k} \rangle \langle \Psi_{j,l}|$, where
\begin{eqnarray}\label{ent1}
|\Psi_{i,k}\rangle &=& \frac{1}{\sqrt{2}} \left[ |1,0\rangle_S|0,1\rangle_T|\psi_i\rangle_{ES}|\psi_k\rangle_{ET} \right. \nonumber \\
&+& \exp(i\theta) \left. |0,1\rangle_S|1,0\rangle_T|\psi_k\rangle_{ES}|\psi_i\rangle_{ET}\right].
\end{eqnarray}
After tracing out the environmental states we gain a partially entangled state of qubits
$\rho_{ST} = \frac{1}{2} \{ |1,0 \rangle_S \langle 1,0| \otimes |0,1\rangle_T \langle 0,1| + |0,1 \rangle_S \langle 0,1| \otimes |1,0 \rangle_T \langle 1,0|
+ [ D \exp(i \theta) |0,1 \rangle_S \langle 1,0| \otimes |1,0 \rangle_T \langle 0,1| + \mbox{h.c.} ] \}$, where
$ D = \sum_{i,j} c_{ij,ji} = \Tr \left[ F \rho_{E,ST} \right] $
is a phase damping (decoherence) parameter which is equivalent to Eq.(\ref{D-def}).
To complete the transfer, we measure qubit $S$ by the projective measurement, $\Pi_{S,\pm}=|\pm\rangle_S\langle\pm|$, where $|\pm\rangle_S=\frac{1}{\sqrt{2}}(|0,1\rangle_S\pm |1,0\rangle_S)$. Then the target qubit is transferred to two possible conditional states $\rho_T^\pm = \frac{1}{2} \{ |0,1 \rangle_T \langle 0,1| + |1,0 \rangle_T \langle 1,0| \pm [ D \exp(i\theta) |0,1 \rangle_T \langle 1,0|
+ \mbox{h.c.} ] \}$, respectively to the measurement result. By conditional application of $\pi$-flip,
$|1,0 \rangle_T \rightarrow -|1,0 \rangle_T$ whereas $|0,1 \rangle_T$ remaining unchanged, we reach state
\begin{equation}\label{state_out}
  \rho_T \equiv \rho_T^+ = \frac{1+D}{2} \ket{\Psi}_S \bra{\Psi} +
  \frac{1-D}{2} \ket{\Psi^\perp}_S \bra{\Psi^\perp},
\end{equation}
where $\ket{\Psi^\perp}_S = \frac{1}{\sqrt{2}}[|0,1\rangle_S - \exp(i\theta)|1,0\rangle_S]$ is the orthogonal complement to $\ket{\Psi}_S$.
State (\ref{state_out}) corresponds to the original qubit state, $|\Psi\rangle_S$, disturbed by decoherence, with its off-diagonal elements (in the computation basis) reduced by factor $D$.
The sign of $D$ does not play a principal role. If it is priori known it can be simply compensated by the same feed-forward mechanism. Thus the quality of this basic quantum transfer can be measured by $\left| \Tr \left[ F \rho_{E,ST} \right] \right|$.
However, perfect transfer with $|D|=1$ can correspond to three very different environmental states $\rho_{E,ST}$: (i) Product of pure perfectly overlapping single-particle states ($D=1$), (ii) Symmetric maximally entangled state ($D=1$), or (iii) Anti-symmetric maximally entangled state ($D=-1$). Although $|D|=1$ for all these cases, it varies differently when these states undergo a local operation $U_{ES}\otimes V_{ET}$.

From the presented point of view, particles $S$ and $T$ in state $|1\rangle_S\langle 1|\otimes|1\rangle_T\langle 1|\otimes\rho_{E,ST}$ can be called {\em effectively indistinguishable} if any qubit carried by particle $S$ (encoded into the degrees of freedom which are supposed to be under experimentalist control like spatial or polarization modes) can be perfectly transferred to a qubit carried by particle $T$ and vice versa. The above described operation transfers a quantum state between two particles similarly as quantum teleportation does. However, entanglement is not used as a resource here, so it is not limited by the amount of entanglement (opposite to Ref.~\cite{ent_wit_exp}). What is important is indistinguishability of particles $S$ and $T$ and it can be well characterized by parameter $|D|$.

Parameter $D$ plays the same role also in the complementary task to quantum state transfer -- quantum erasure. Due to the symmetry of state $\rho_{ST}$, to concentrate phase information back to qubit $S$ we can apply the same type of measurement (but now on qubit $T$) and feed-forward strategy (on qubit $S$). Since the reconstructed state of qubit $S$ has the same structure as above with $D$ given by Eq.~(\ref{D-def}), $|D|$ represents the same upper limit on the quality of quantum erasure.

\section{Direct measurement of indistinguishability}

To experimentally determine $|D|$ one can advantageously use the standard HOM-type experiment \cite{HOM}, similarly as it has been used for the measurement of an overlap \cite{Filip2002,Hendrych2003}. Particles enter a balanced unitary mixer which symmetrically transmits and reflects them between modes $S$ and $T$. If basis state $|1 \rangle_S |1 \rangle_T | \psi_i \rangle_{ES} | \psi_k \rangle_{ET}$ is in the input then the corresponding post-selected output state (one particle in $S$ and the other one in $T$ mode) is proportional to $\frac{1}{2} |1 \rangle_S |1 \rangle_T \left( |\psi_i \rangle_{ES} |\psi_k \rangle_{ET} + |\psi_k \rangle_{ES} |\psi_i \rangle_{ET} \right)$. For input state $|1\rangle_S\langle 1|\otimes|1\rangle_T\langle 1|\otimes\rho_{E,ST}$ the probability of  coincidence detection, $P_C = \frac{1}{2} \left( 1 - \sum_{i,j} c_{ij,ji} \right) = \frac{1}{2} ( 1 - D )$, is directly proportional to parameter $D$.

Experimental quantum information processing and transfer often uses photonic qubits \cite{Kok2007,Gisin2002} encoded into photons generated by spontaneous parametric down-conversion (SPDC). These photons represent a typical example of qubit carriers with internal degrees of freedom which may exhibit complex behavior \cite{Pan2012,Torr2011}.
State of two photons created by SPDC, which propagate in well defined spatial and polarization modes, can be described by the following formula:
\begin{equation}\label{stateSPDC}
 \ket{\psi} = \int_{\omega} \phi(\omega) \phi(\omega_{0}-\omega) \, e^{-i \, \omega \, \Delta t} \, \ket{\omega}_a \ket{\omega_{0}-\omega}_b,
\end{equation}
where $\ket{\omega}_x$ represents a single photon at spatial mode $x$ and frequency $\omega$, $\Delta t$ denotes time delay in mode $a$, and $\phi(\omega)$ is a spectral amplitude function. This function is mainly determined by the employed spectral filter. In our case the filter is the same both for mode $a$ and $b$. Let us suppose this state enters a balanced (50:50) beam splitter (BS) with input modes $a,b$ and output modes $c,d$. The input state is transformed to the output one by a unitary transformation, $\ket{\varphi} = U_{BS} \ket{\psi}$, and the creation operators of input modes can be expressed by the creation operators of output modes as follows
\begin{eqnarray*}
 a^\dag(\omega) &=& \frac{1}{\sqrt{2}} \left[ i \, c^\dag(\omega) + d^\dag(\omega) \right],\\
 b^\dag(\omega) &=& \frac{1}{\sqrt{2}} \left[ c^\dag(\omega) + i \, d^\dag(\omega) \right].
\end{eqnarray*}
Let us further suppose that behind the BS we make a coincident measurement. Coincidence rate can be calculated as
$
R(\Delta t) \propto \int_{t_1} \int_{t_2} \bra{\varphi} \, E^{(-)}_c(t_1) E^{(-)}_d(t_2) E^{(+)}_d(t_2) E^{(+)}_c(t_1) \, \ket{\varphi},
$
where $E^{(+)}_x(t) \propto \int_{\omega} x(\omega) \, e^{-i \, \omega \, t}$ is a positive-frequency part of an electric field operator with $x$ being an annihilation operator and $E^{(-)}_x(t) = [ E^{(+)}_x(t) ]^\dag$  \cite{MandelWolf}. Here we assume the coincidence window to be infinitely large. In practice it is about 1\,ns but the two photons arrive in the interval of about 100\,fs which is much shorter.
In this notation the flip operator can be expressed as
$
F = \int_{\omega_{1}} \int_{\omega_{2}} \ket{\omega_{1}}_c \ket{\omega_{2}}_d \, \bra{\omega_{2}}_c \bra{\omega_{1}}_d.
$
It can be shown by a straightforward calculation that
\begin{equation}\label{eqR}
R(\Delta t) \propto 1 - \Tr \left[ F \, \ketbra{\varphi} \right]
= 1 - D
\end{equation}
with $\Tr(X) = \int_{\omega_{1}} \int_{\omega_{2}} \bra{\omega_{1}}_c \bra{\omega_{2}}_d \, X \, \ket{\omega_{1}}_c \ket{\omega_{2}}_d$.
For entangled input state (\ref{stateSPDC}) we obtain
$$
\Tr \left[ F \, \ketbra{\varphi} \right] \propto
\int_{\omega} \left| \phi \left( \frac{\omega_{0} + \omega}{2} \right) \right|^2 \,
\left| \phi \left( \frac{\omega_{0} - \omega}{2} \right) \right|^2 e^{i \, \omega \, \Delta t}.
$$
If $\phi(\omega)$ is a rectangular function of width $v$ and central frequency $\omega_{0}/2$ then $D = \Tr \left[ F \, \ketbra{\varphi} \right] = \sinc(\Delta t \, v)$.

Clearly, in such an experimental situation the role of the ``environment'' is played by the frequency degrees of freedom (in our experiment qubits are encoded into spatial modes). Parameter $D$ can really be measured only by means of a beam splitter and coincidence detection and it can be varied by changing delay $\Delta t$ between the two photons. Its negative values correspond to partially entangled states containing vectors from anti-symmetric subspace.

%%%%%%%%%%%%%%%%%%%%%%%%%%%%%%%%%%%%%%%%%%%%%%%%%%
%%%%%%%%%%%%%%%%%%%%%%%%%%%%%%%%%%%%%%%%%%%%%%%%%%

\section{Experimental setup}

Our setup is depicted in Fig.~\ref{fig-scheme}. Photon pairs are created by collinear frequency-degenerate type-II SPDC in a beta barium borate crystal pumped at 405\,nm. Both photons pass through the same band-pass interference filter of \emph{approximately} rectangular shape with central frequency 810\,nm and spectral width (FWHM) 2.7\,nm.
Then they are separated by a polarizing beam splitter and coupled into single-mode fibers. One of the photons is retarded by $\Delta t$ in a delay line (DL) with adjustable length. By means of polarization controllers the both photons are set to have the same polarizations.
Qubit states are encoded into spatial modes of individual photons. Each basis state corresponds to a single photon in one, $\ket{0,1}$, or in the other, $\ket{1, 0}$, of two optical fibers. Initial equatorial states of both qubits are prepared using fiber couplers (FC and VRC) with splitting ratio 50:50 and integrated electro-optical phase modulators (PM).

In the experiment with the qubit-state transfer, the source qubit, in ``unknown'' equatorial state (\ref{state_S}), was represented by qubit I and the target qubit, in state (\ref{state_T}), was represented by qubit II. In the quantum erasure experiment, the source was qubit II, while the target was qubit I.

The key part of our device is the swap of two rails between qubits I and II followed by measurement on qubit I. This measurement is performed in basis $\frac{1}{\sqrt{2}}(|0,1\rangle \pm |1,0\rangle)$ using a fiber coupler with fixed splitting ratio 50:50 and two single photon detectors (silicon avalanche photodiodes).
When detector $\mathrm{D_{a1}}$ clicks, phase correction $\pi$ is applied on qubit II by means of electronic feed forward \cite{mik12}. Feed forward uses a direct signal from detector $\mathrm{D_{a1}}$ (5\,V pulse). The signal is modified by a passive voltage divider to circa 1.5\,V and then it is lead to a lithium-niobate phase modulator (1.5\,V corresponds to the phase shift of $\pi$).
States of output qubit II are characterized by quantum tomography. Different measurement bases are set by a phase modulator and variable ratio coupler (VRC). Photons are counted by detectors $\mathrm{D_{d0}}$ and $\mathrm{D_{d1}}$. Small differences in detector efficiencies are corrected numerically in the data sets.

The whole experimental setup consist of two interconnected Mach-Zehnder interferometers. Lengths of their arms are balanced by motorized air gaps (not shown in the figure). To reduce a phase drift caused by environmental influences (temperature fluctuations etc.) the whole setup is covered and also actively stabilized. After each 3\,s period of measurement the phase drifts are determined and compensated by adding a proper correcting voltage on phase modulators.
The HOM dip, which we use to characterize the properties of input photons, is measured at the last VRC \cite{HOM}.

%%%%%%%%%%%%%%%%%%%%%%%%%%%%%%%%%%%%%%%%%%%%%%%%%%
\begin{figure}
  \begin{center}
   %\smallskip
  \resizebox{\hsize}{!}{\includegraphics*{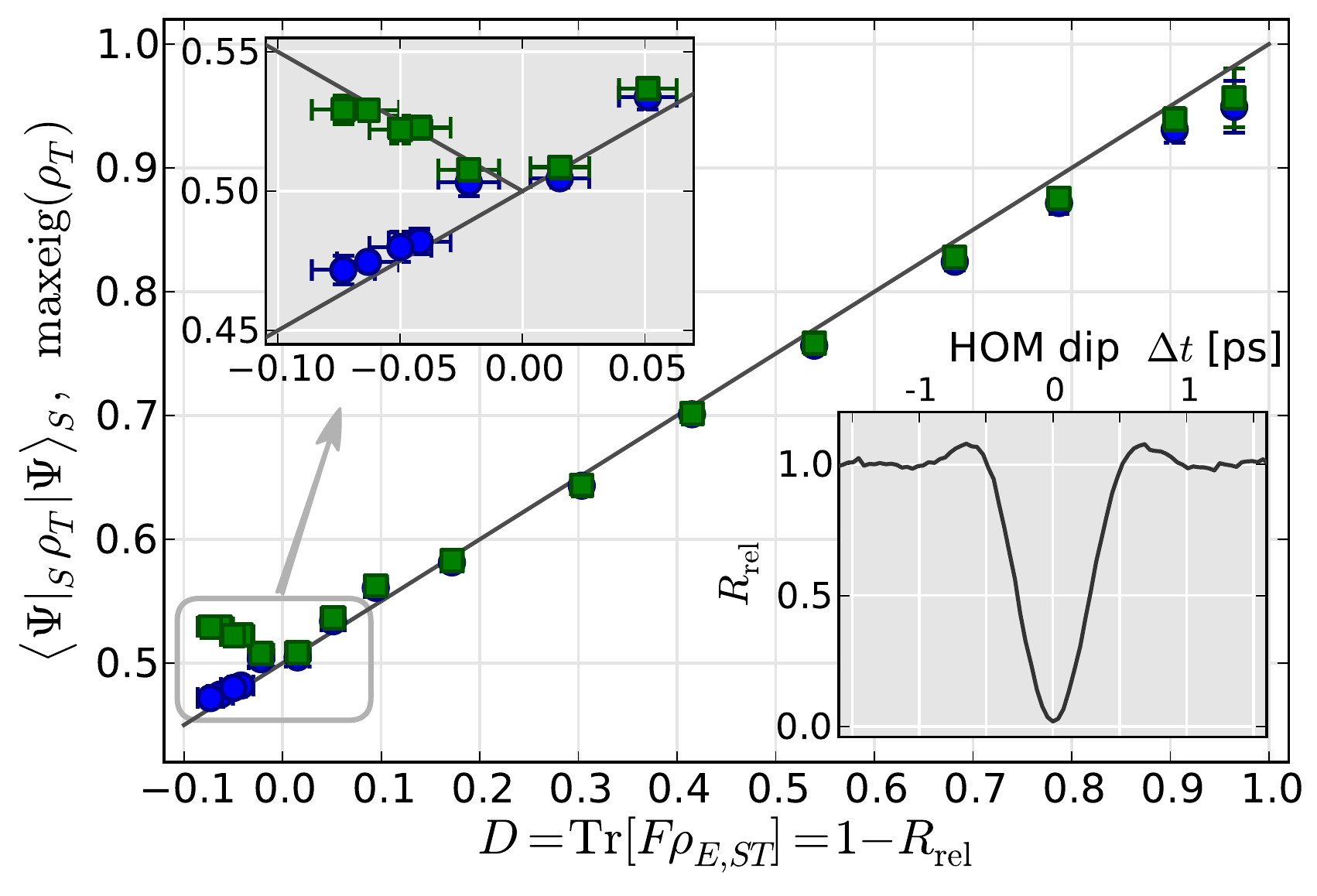}}
   %\smallskip
  \end{center}
  \caption{(Color online) Dependence of the quality of qubit-state transfer on parameter $D$. Blue circles denote the overlap of output and input states, $\bra{\Psi}_S \rho_T \ket{\Psi}_S$. Green squares denote maximal eigenvalues of output states $\rho_T$. Straight lines are theoretical predictions. The upper left inset magnifies the area where $D$ is close to zero. The lower right inset shows the measured Hong-Ou-Mandel dip, $R_\mathrm{rel}$ denotes relative (normalized) coincidence rate.}
  \label{fig-results}
\end{figure}
%%%%%%%%%%%%%%%%%%%%%%%%%%%%%%%%%%%%%%%%%%%%%%%%%%

\section{Results}

We have tested both the qubit-state transfer and quantum erasing. But here we will discuss only the qubit-state transfer because the results of quantum erasing are quite similar (as can be expected from the symmetry of the tasks).
The target, qubit II, was prepared in state (\ref{state_T}) and the source, qubit I, was prepared in state (\ref{state_S}) with phase $\theta = 0, 30, 60, 90, 120, 150, 180$ degrees, in sequence. At the output we have made measurement on the target qubit (qubit II) in three different bases:
$\{ |0,1\rangle, |1,0\rangle \}, \{ \frac{1}{\sqrt{2}}(|0,1\rangle \pm |1,0\rangle) \}$, and
$\{ \frac{1}{\sqrt{2}}(|0,1\rangle \pm i |1,0\rangle) \}$.
Each measurement consisted of 15 three-second measurement intervals interlaced by active stabilization. The results were used to reconstruct output density matrices, $\rho_T^\mathrm{rec}$, by means of maximum-likelihood quantum tomography \cite{par04,hra97,jez03}. Reconstruction of density matrices has enabled us to calculate various quantities including purity, Uhlmann fidelity (with respect to theoretical output states), eigenvalues, and overlap with corresponding input states.

Each such a measurement set was repeated 16 times with different delays $\Delta t$ between the input photons (corresponding to different positions in HOM dip). For each $\Delta t$ we have also evaluated parameter $D=1-R_\mathrm{rel}$. It was obtained from the coincidence rate between detectors $\mathrm{D_{d0}}$ and $\mathrm{D_{d1}}$, $R(\Delta t)$, normalized with respect to the coincidence rate measured separately in the position far from the dip, $R(2\,\mathrm{ps})$, i.e., $R_\mathrm{rel} = R(\Delta t)/R(2\,\mathrm{ps})$. Negative values of $D$ correspond to the positions in the raised ``shoulders'' of the HOM dip. They reveal that the ``environments'' of our photons are entangled. The values of $\bra{\Psi}_S \rho_T \ket{\Psi}_S$ lower than $0.5$ mean that the roles of states $\ket{\Psi}_S$ and $\ket{\Psi^\perp}_S$ were swapped (see Eq.~\ref{state_out}).

According to the theory, overlap $\bra{\Psi}_S \rho_T \ket{\Psi}_S = \frac{1+D}{2}$ and eigenvalues of $\rho_T$ are $\frac{1+D}{2}$ and $\frac{1-D}{2}$, see Eq.~(\ref{state_out}). Fig.~\ref{fig-results} shows the overlap and the maximal eigenvalue as functions of parameter $D$. Each point represents an average over all 7 phases.
Vertical error bars visualize standard deviations obtained from the ensembles of measurements with different phases. Due to various experimental imperfections (phase fluctuations, drift of splitting ratios, etc.) they are greater than standard deviations calculated purely from Poissonian photo-count distribution. But on the graph they are mostly smaller than the size of the symbols. Horizontal error bars reflect (Poissonian) statistical fluctuations of coincidence rates $R(\Delta t)$ and $R(2\,\mathrm{ps})$. Average output state fidelity, $\left[ \Tr \left( \sqrt{ \sqrt{\rho_T} \rho_T^\mathrm{rec} \sqrt{\rho_T} }  \right) \right]^2$, (averaged over all phases and all delays) was $99.2 \pm 0.8 \%$.
The measured HOM dip is shown in the lower right inset of Fig.~\ref{fig-results}.
Relative measurement error was less than 6\,\% in its minimum and less than 2\,\% for maximal values. Dip visibility was $96.4 \pm 0.4 \%$.

\section{Conclusions}

We can conclude that effective quantum indistinguishability as a key resource for quantum information processing can be quantified by a directly measurable parameter $\left| \Tr \left[ F \rho_{E,ST} \right] \right|$ for any state $\rho_{E,ST}$ of two particles. We have demonstrated that this parameter represents a bound on the quality of real-world implementation of quantum transfer protocols.
If other resources, like quantum entanglement, are required then the impact of their imperfections is combined with the effect of (in)distinguishability in a nontrivial way.
Their coexistence is a subject of a current investigation.

\begin{acknowledgements}
This work was supported by Palack\'{y} University (PrF-2012-019). R.\,F. acknowledges support from the Czech Science Foundation (P205/12/0577).
\end{acknowledgements}

\end{document}